\documentclass{epl}

\title{Loading of a Bose-Einstein condensate 
in the boson--accumulation regime}

\author{F. Floegel, L. Santos and M. Lewenstein}

\institute{Institut f\"ur Theoretische Physik, Universit\"at Hannover,
 D-30167 Hannover,Germany}


\begin{document}

\maketitle

\begin{abstract}
We study the optical loading of a trapped Bose-Einstein condensate
by spontaneous emission of atoms in  excited electronic state in the 
Boson-Accumulation Regime. 
We generalize the previous simplified analysis of 
ref. [{\it Phys. Rev.} A {\bf 53}, 2466 (1996)], 
to a 3D case in which more than one trap level of the excited state trap 
is considered.
By solving the corresponding quantum many--body master equation, 
we demonstrate that 
also for this general situation the photon reabsorption 
can help to increase the condensate fraction. Such effect 
could be employed to realize a continuous atom laser,  
and to overcome condensate losses.
\end{abstract}

During the last years the experimental realization of 
Bose--Einstein condensation (BEC)
 in trapped weakly interacting gases \cite{BEC}
has stimulated an enormous interest \cite{Varenna}. 
Among the results related to BEC, specially remarkable is
the realization of an {\em Atom
Laser} \cite{rf,raman,kasev98,haensch}.  As a
coherent source of matter waves, the
atom laser will lead to new applications in atom optics. 
Several groups have 
demonstrated atom lasers using pulsed \cite{rf,raman} or continous 
\cite{haensch} outcoupling from the BEC, either by  employing rf fields 
\cite{rf}, or 
Raman pulses \cite{raman}. However, the continuous outcoupling 
is just a half way towards a cw atom laser. The continuous loading
of the condensate still remains to be realized.
Like in the developement of light lasers the
availability of cw atom lasers would open the way to ``high power''
and precision applications.

Two different physical mechanisms could provide continuous
pumping of atoms into a condensate. On one hand, this can be achieved 
by collisional mechanisms \cite{Castin}, in which two non--condensed atoms 
collide, one being pumped into the
condensate, whereas the other being evaporated. On the other hand, the optical pumping 
into a BEC via spontaneous emission has also been
proposed \cite{Martin}. If the reservoir could be filled
in a (quasi-) continuous way  by laser cooling techniques, one
would benefit from the large cooling efficiency of laser cooling
compared to evaporative cooling, allowing for a considerable
increase in atomic flux produced by an atom laser. For the
latter, it is crucial that the spontaneously emitted photons
cannot be reabsorbed, since otherwise heating is introduced
in the system, and BEC can be neither achieved, nor maintained
\cite {Dalibard}. It has been shown that in the 
so-called Festina Lente regime, in which 
the spontaneous emission rate $\gamma$ is 
smaller than the trap frequency $\omega$, 
the heating due to the reabsorption
processes is suppressed  \cite{Festina}. 
However, due to the
slow time constants in this approach the cooling efficiency is
greatly reduced. Recently, it has been also reported 
\cite{BRE} that when an atom possesses an accessible 
three level $\Lambda$
scheme, where one of the atomic transitions decays much faster
than the other, the reabsorptions in the slow transition 
can be largely suppressed, without the time limitations of Festina Lente. 

In this Letter we study the loading of a BEC, which is formed in a trapped 
electronic  ground--state $|g\rangle$, via spontaneous decay of atoms from an 
also  trapped internal excited state $|e\rangle$.
In the regime $N_0 \gg a,N-N_0$ (where $a$ is
the effective number of levels other than the condensed one to
which the excited atoms may decay, $N$ is the total number of atoms 
in the $|g\rangle$ trap, and $N_0$ is the number of condensed ones)
the atom will
decay into the condensate with high probability, due to the
bosonic enhancement. Such regime 
has been named {\it Boson--Accumulation Regime} (BAR) \cite{BAR}. 
Note, however, that even in the BAR, in the absense of reabsorption 
simple arguments imply that the mean
proportion of atoms in the condensate after the decay
decreases \cite{BAR}. 
If the reabsorptions are present, each subsequent decay would lead to   
an even larger decrease of the condensate proportion; 
thus for an optically thick sample (where many reabsorptions
take place) the number of condensed atoms would be reduced
dramatically. However, these  arguments are not rigorously valid in general. 
In ref.\ \cite{BAR} it was reported that a fully quantum treatment
shows that the reabsorption processes can, under certain conditions, 
help to increase the proportion of atoms in the BEC.
This counterintuitive result was explained by means of 
an interference effect between different
paths (which include reabsorptions) that lead to the
same final states. This effect offers an interesting mechanism 
of continuous refilling and loading of a condensate. 

The treatment of ref.\ \cite{BAR} was based on an extremely
simplified 1D model in which all the ground states were taken into
account, but in which there was a single excited level. 
In addition, the calculations of 
 ref.\ \cite{BAR} did not consider the different Frank-Condon 
factors for the different possible decays.
In this Letter, we analyze the BAR regime in a much more general 
model. We consider a 3D trap, and the possibility of 
more than one excited trap level. 
In addition, the Frank-Condon factors 
for the different decays are explicitely included in the calculations.
We demonstrate that even for this much more complicated situation,
under some conditions the reabsorption still helps to 
load atoms  into the condensate.

Let us consider a set of bosonic atoms with two internal levels $|g\rangle$ and 
$|e\rangle$ confined in a dipole harmonic trap, 
which for simplicity is considered isotropic, with frequency $\omega$.
We denote as $N_m$ the population of the $m$-th level of the $|g\rangle$ trap, where $m\equiv (m_x,m_y,m_z)$, and 
assume that the ground state trap verifies the BAR conditions.
We analyze the situation in which a single atom in some state
$|e,l\rangle$ ($l\equiv (l_x,l_y,l_z)$)
decays via spontaneous emission into the $|g\rangle$ trap, 
producing a photon which can  eventually be reabsorbed by another $|g\rangle$ atom
(in particular by a condensed one), 
which in turn can later decay again, and so on. At some finite time, the scattered photon 
is no more reabsorbed and leaves the system. It is our aim to investigate how the atom distribution in 
the $|g\rangle$ trap changes during 
this process.
 
For simplicity we do not consider the collisional mean--field effects. For typical $s$--wave scattering lengths
this implies that in principle our results could be obtained
for few atoms only (since we consider relatively small traps); one should point out 
, however, that although our calculations  are 
limited to small traps, the BAR effect should be present also for larger 
ones, and in the presence of 
collisions. Although the main purpose of this Letter is to discuss a fundamental quantum effect and 
to present the methodology needed to study it, we must also point out that  
recent experiments \cite{MIT} have shown that the $s$-wave scattering length 
can 
be modified by using Feshbach resonances. In particular, the regime of 
quasi--ideal 
gas can be experimentally achieved, in which the mean--field energy can be considered  
smaller than the trap energy. In such regime, the collisions introduce just a 
thermalization mechanism \cite{collisions}, and two and three--body collisional losses are 
almost  absent \cite{Rb85Feshbach}. For such modified scattering length our results are valid for much 
larger $N$.

\begin{figure}[ht] 
\threeimages[scale=0.8]{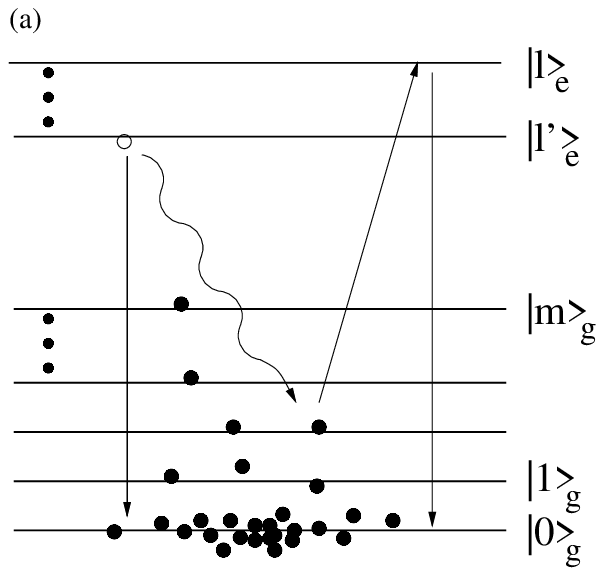}{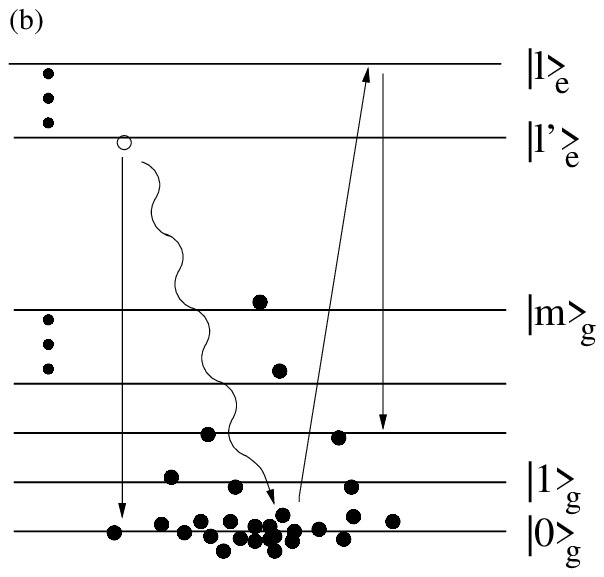}{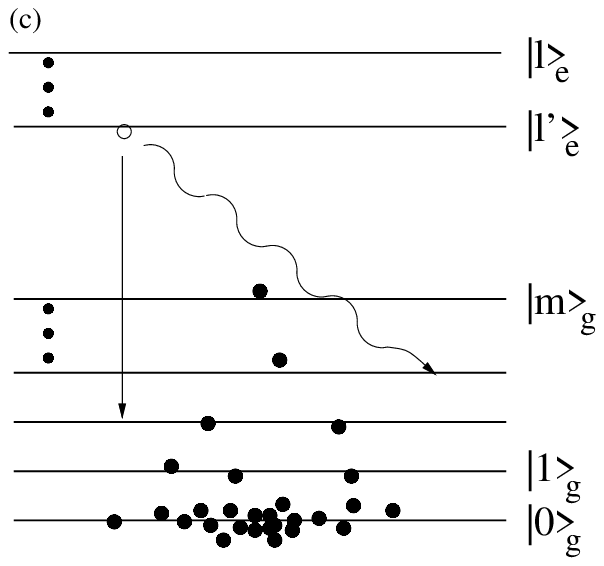}
\caption{Schematic representation of the different processes appearing in eqs.\
(\ref{PN0+2})
and (\ref{PN0s}). Straight lines denote atomic transitions, whereas waved lines
represent
the photons produced in the decays. Note the possibility of reabsorption 
in figs. (a) and (b).}
\label{fig:1}  
\end{figure}

Starting from the Hamiltonian which describes the bosons
interacting with the quantized electromagnetic field, and using
standard techniques, one can derive the master equation (ME) for the
reduced density operator for the atomic degrees of freedom
\cite{Gardiner}, which assuming $\hbar=1$ becomes
\begin{equation}
\label{3MEgen}
\dot \rho = - i H_{eff} \rho + i \rho H_{eff}^\dagger + J\rho,
\end{equation}
where 
\begin{equation}
H_{eff} = \sum_{l=0}^\infty (\omega_l^e+\omega_0-i\frac{\Gamma}{2}) e^\dagger_l e_l 
+ \sum_{k=0}^\infty \omega_k^g g^\dagger_k g_k   
-i \frac{\Gamma}{2}  \sum_{l,l^\prime}^\infty \sum_{m,m^\prime}^\infty 
\alpha_{lmm^{\prime} l^{\prime}} g^\dagger_{m^\prime} g_m 
e^\dagger_l e_{l^\prime},
\end{equation}
is a non--hermitian effective Hamiltonian, and
\begin{equation}
J\rho = \Gamma  \sum_{l,l^\prime}^\infty \sum_{m,m^\prime}^\infty 
 \alpha_{lmm^{\prime} l^{\prime}}^r 
g^\dagger_{m^\prime}  e_{l^\prime}  \rho e^\dagger_l g_m.
\end{equation}
is the jump operator. Here, 
$e_l$ ($g_m$)  is the annihilation operator for atoms in
the $l$--th ($m$--th) excited (ground) level,
$\omega_l^e$ ($\omega_m^g$) is the energy corresponding to such state, and
$\Gamma$ is the spontaneous emission rate from the
excited level. The complex coefficients $\alpha_{lmm^{\prime} l^{\prime}}$ 
(whose explicit form can be found in \cite{BAR}) 
are related to the Frank-Condon factors for the different excited--ground transitions.
We denote in the following $\eta_{lm}$ the Franck-Condon factor
for the transition between  the $l$--th excited level
and the $m$-th ground  level. The real part of the $\alpha_{lmm^{\prime} l^{\prime}}$
coefficient is denoted $\alpha_{lmm^{\prime} l^{\prime}}^r$.  
We consider an initial situation in which $N_m$ atoms occupy the 
$m$-th state of the ground--state trap, 
and a single excited atom is placed (with some probability given by a 
thermal distribution) in a state $j$ of its corresponding trap.
We denote the initial state as $|\psi_0\rangle$, and therefore 
the initial density matrix is defined as
$\rho_N=|\psi_0\rangle\langle\psi_0|$. 
After a photon is released without further reabsorptions,
we obtain the formal solution $\rho_{N+1}$ of the ME (\ref{3MEgen}), 
and calculate the probability to obtain a particular final state 
$|\psi_f\rangle$ with $N'_m$ atoms in the $m$-th $|g\rangle$ trap level 
(and no atom in the excited state trap). 
After expanding into powers of the small parameter $N_0^{-1/2}$, and neglecting 
terms of order ${\cal O}(N_0^{-3/2})$, such probability takes the form
\begin{eqnarray}
&&\langle\psi_f|\rho_{N+1}|\psi_f\rangle=\frac{2}{N_0}\int_0^{\infty}dt
\int\frac{d\Omega}{4\pi}  \left | \sum_l\eta_{l0}^{\ast}|\langle\psi_f|g_0^\dagger e_l
A_1|\psi_0\rangle + \sum_{l,m\not=0}\eta_{lm}^{\ast}\langle\psi_f|g_m^\dagger e_l
A_0|\psi_0\rangle \right |^2
\end{eqnarray}
with $A_0(t)=\exp[-iH_{eff}^{(0)}t]$ and 
$A_1=-i\int_0^t d\tau \exp[-iH_{eff}^{(0)}(t-\tau)] H_{eff}^{(1)} \exp[-iH_{eff}^{(0)}\tau]$ 
being the terms of $\exp[-iH_{eff}t]$ of order $0$ and $1$ in $1/N_0^{1/2}$, where
\begin{eqnarray}
H_{eff}^{(0)} = 
  \omega_0^g g^\dagger_0 g_0
-i \frac{\Gamma}{2}  \sum_{l,l^\prime}^\infty 
\alpha_{l00^{\prime} l^{\prime}} g^\dagger_{0} g_0 
e^\dagger_l e_{l^\prime},
\end{eqnarray}
and
\begin{eqnarray}
H_{eff}^{(1)} =
-i \frac{\Gamma}{2}  \sum_{l,l^\prime}^\infty \sum_{m \not = 0}^\infty 
\{ \alpha_{lm0 l^{\prime}} g^\dagger_0 g_m +
   \alpha_{l0m l^{\prime}} g^\dagger_{m} g_0 
\}
e^\dagger_l e_{l^\prime},
\end{eqnarray}
are respectively the terms of $H_{eff}$ of zeroth and first order in 
$1/N_0^{1/2}$. Therefore, beginning from an initial relative number of condensate particles $n=N_0/N$, 
up to order ${\cal O}(1/N_0^2)$, the new relative number of 
condensed atoms after the last decay becomes 
\begin{equation}
n^\prime = \frac{N_0+1+(P_{N_0+2}-P_{N_0})}{N+1} ,
\end{equation}
where $P_{N_0+2}$ ($P_{N_0}$) is the probability 
to have after the process $N'_0=N_0+2$ ($N_0$). 
Let $|\psi_{f}^{(2,s)}\rangle$ be the final state with $N'_0=N_0+2$, 
$N'_s=N_s-1$, and $N'_{j\not= 0,2}=N_j$.
The probability to decay into such a state is given by:
\begin{equation}
P_{N_0+2}^{s}=\langle\psi_f^{(2,s)}|\rho_{N+1}|\psi_f^{(2,s)}\rangle = \frac{2}{N_0}\int_0^\infty dt
\int\frac{\Omega}{4\pi}
\left | \sum_{l,m\not=0}\eta_{lm}^{\ast}\langle\psi_f^{(2,s)}|g_m^\dagger e_l
A_0(t)|\psi_0\rangle \right |^2.
\label{PN0+2}
\end{equation}
The corresponding process is schematically represented in 
fig.\ \ref{fig:1}(a).
Let us also define 
$|\psi_f^{(0,s)}\rangle$ as the final state with $N'_0=N_0$, 
$N'_s=N_s+1$ and $N'_{j\not= 0,2}=N_j$.
The probability to decay into this state takes the form:
\begin{eqnarray}
P_{N_0}^{s}&=&\langle\psi_f^{(0,s)}|\rho_{N+1}|\psi_f^{(0,s)}\rangle\nonumber \\
&=&
\frac{2}{N_0}\int_0^\infty dt
\int\frac{\Omega}{4\pi} \left \{  
\left | \sum_l\eta_{l0}^{\ast} \langle\psi_f^{(0,s)} |g_0^\dagger e_l
A_1(t)|\psi_0\rangle\right |^2 + 
\left | \sum_{l,m\not=0}\eta_{lm}^{\ast}\langle\psi_f^{(0,s)}|g_m^\dagger e_l
A_0(t)|\psi_0\rangle \right |^2 \right. 
 \nonumber \\
&& \left. + 2\Re\left\{ \sum_l\eta_{l0}^{\ast}|\langle\psi_f^{(0,s)}|g_0^\dagger e_l
A_1(t)|\psi_0\rangle \left (\sum_{l,m\not=0}\eta_{lm}^{\ast}\langle\psi_f^{(0,s)}|g_m^\dagger e_l
A_0(t)|\psi_0\rangle \right )^\ast \right \} \right \}.
\label{PN0s}
\end{eqnarray}
where $\Re$ denotes the real part. The first and second terms in eq.\ (\ref{PN0s}) are respectively depicted 
in fig.\ \ref{fig:1}(b) and (c), whereas the last term corresponds to the intereference between the 
energetically equivalent paths which are shown in that figures.
From eqs.\ (\ref{PN0+2}),(\ref{PN0s}) one obtains 
$P_{N_0+2}=\sum_s <P_{N_0 + 2}^{s}>$, 
and $P_{N_0}=\sum_s <P_{N_0}^{s}>$, where $<>$ denotes 
the average of the possible initial ground--state populations 
following the corresponding Bose--Einstein distribution.
We have numerically calculated the probabilities $P_{N_0}$ and 
$P_{N_0+2}$. From eqs.\ (\ref{PN0+2}),(\ref{PN0s}) it becomes clear that such calculation 
requires to take into account all the possible paths 
connecting a particular initial and final state
which eventually involves an arbitrary number of emission-reabsorption
cycles. This fact by itself makes the calculation extremely demanding. 
In addition, several important technical difficulties appear.
First, we must note that in order to calculate the coefficients 
$A_0$ and $A_1$, one has to evaluate 
the exponential $\exp(-i\hat M t)$, where 
$\hat M=\sum_{l,l^\prime}\alpha_{l00l^\prime} e_l^\dagger e_{l^\prime}$
is not an Hermitian matrix. Therefore, 
it becomes neccesary to introduce a biorthogonal set of operators
$f_R^{\dagger}(k)=\sum_l v_l^R(k)e_l^\dagger$,  
and $f_L(k)=\sum_l {\bar v}_l^L(k)e_l$, 
where $v_l^R(k)$ and ${\bar v}_l^L(k)$
are the right and the left (complex conjugate) eigenvectors of $\hat M$.
Therefore, such eigenvectors, and eigenvalues have to be calculated.
Second, we note that the $\alpha$ coefficients constitute a $12$ dimensional 
tensor, 
and that in order to evaluate every component, it is neccesary to
calculate the Franck--Condon factors for each possible 
excited--ground transition, and to perform
the non--trivial task of calculating the Cauchy principal part integral 
appearing in the imaginary part of the $\alpha$ coefficients 
\cite{BAR}. Such strong technical difficulties 
make eventually impossible to solve in 3D in a reasonable computational time
systems with more than $4$ excited--state shells ($20$ levels), and therefore 
we have been constrained by such limit. 

\begin{figure}[ht] 
\onefigure[scale=1.1]{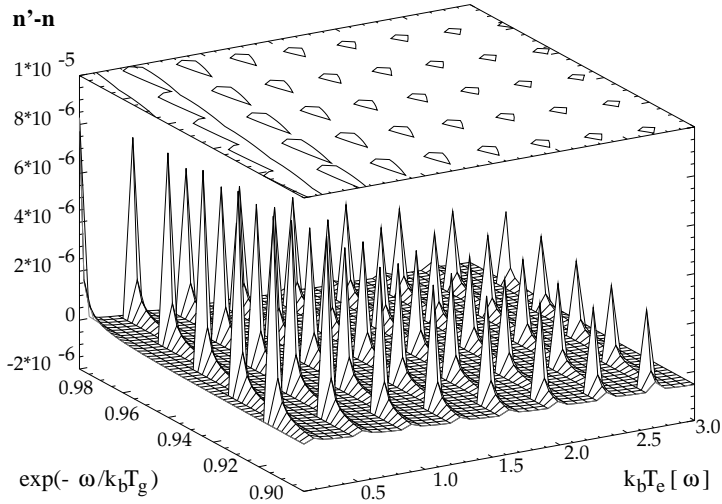}
\caption{
Change of the condensate fraction after a decay process
as a function of
the temperature $T_e$ of the $|e\rangle$--trap and $T_g$ of the $|g\rangle$--trap, the latter in exponential
scale.
The case of a total number of atoms $N=7\times 10^4$, $10$ energy shells in the $|g\rangle$--trap, $4$ energy shells in the
$|e\rangle$--trap and $\eta^2=2$ is considered. On the top of the figure, the regions enclosed by the contours denote those
temperatures for which the condensate fraction increases during the decay 
process}
\label{fig:2}  
\end{figure}

We have analyzed the case of different number of energy shells of the  $|e\rangle$ and  $|g\rangle$ trap, 
for different temperatures and different number of atoms. For each case, we have calculated the 
probabilites (\ref{PN0+2}),(\ref{PN0s}), in order to evaluate the new relative number of condensed atoms 
$n'$. For the particular case of a single level in the excited state trap, results similar to those obtained 
in ref.\ \cite{BAR} were obtained. Figure \ref{fig:2} shows for different temperatures of the excited and ground trap,
the case of $4$ shells ($20$ levels) in the $|e\rangle$ trap, $10$ shells ($220$ levels) in the $|g\rangle$ trap, 
$N=7\times 10^4$ atoms, and a Lamb--Dicke parameter 
$\eta^2=\omega_r/\omega=2$, where $\omega_r$ is the recoil energy of the scattered photon.
One can observe that for certain temperatures $T_e$ and $T_g$ of the  
$|e\rangle$ and  $|g\rangle$ traps, the change in the relative number of 
condensed particles, $n'-n$ is maximally positive, {\it i.e.} 
at these maxima the reabsorption effects help in the most efficient way to load the condensate.
Away from such maxima, $n'-n$ is positive for low temperature $T_{e}$ of the excited state trap
and not very low temperatures $T_g$ of the $|g\rangle$.
For very low $T_g$, however, the positive processes described by $P_{N0+2}$
tend to vanish, since the number of non--condensed atoms which eventually
could be repumped to the condensate  becomes very small.
On the top of fig. \ref{fig:2}, such regions are those enclosed by the contours.
Out of these regions, $n'-n$ becomes negative,
i.e. the reabsorption tends to decrease the condensation relative number.
In all our calculations the BAR expansion has been proved to be valid, by 
checking the
condition $P_{N_0},P_{N_0+2}\ll 1$.
We have in general observed that the BAR can fail not only for large $T_g$, 
as expected from ref.\ \cite{BAR},
but eventually also, for a given total number of atoms $N$,
for large $T_e$ at the $n'-n$ maxima. The latter is due to the  
large values of the imaginary part of the $\alpha$ coefficients at those peaks 
under such conditions, which 
invalid the expansion performed to obtain eqs.\  (\ref{PN0+2}) and (\ref{PN0s}).  
In the presented example, the condition above is fulfilled
except for the maxima in the region $k_B T_e>\omega$ and $k_B T_g > 50 \omega$.   
As discussed in ref.\ \cite{BAR}, the phenomenon behind the positive effects of 
the reabsorption ($n'-n>0$) cannot be explained by using 
(classical) rate equations, since it is given by the interference between the 
different paths which lead to the same final state.
In particular, it becomes decisive that the interference term in eq.\ (\ref{PN0s})
is always destructive, since the process of fig.~2(b)
includes an additional absorption--emission cycle (which gives
a minus sign in the amplitude, like for a 2$\pi$--laser pulse).  
The consequence of this
interference is that $P_{N_0}$ decreases, which favours that
the excited atom goes to the state $|N_0+1\rangle$, and
therefore it contributes to an increase  of the proportion of
condensed atoms. 

\begin{figure}[ht] 
\onefigure[scale=0.9]{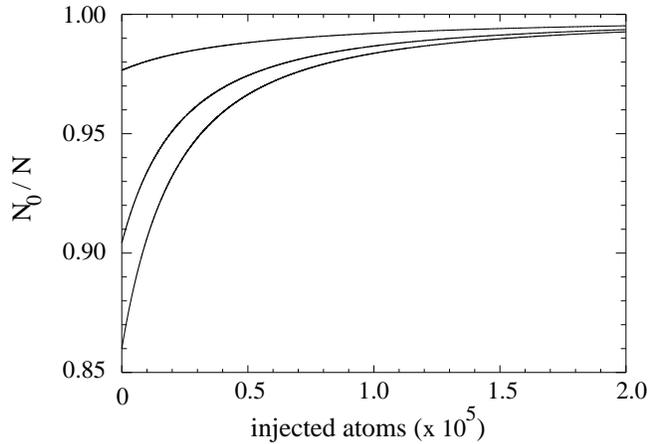}
\caption{
Mean condensate fraction as a function of the number of loaded atoms, for the case of
$10$ energy shells in the $|g\rangle$--trap, $4$ energy shells in the $|e\rangle$--trap and $\eta^2=2$, and
an initial total number of atoms $N=5\times 10^4$.
From top to bottom, the curves represent, respectively, the case of an initial
number of condensed particles $N_0=49632$, $48828$ and $45211$.}
\label{fig:3}  
\end{figure}

We have numerically simulated the pumping of atoms into a BEC in the $|g\rangle$ trap, under the BAR conditions. In our 
calculations, we have assumed that the collisions act on a shorter time scale than the spontaneous optical pumping. 
In that case, collisions will provide a fast thermalization mechanism of the $|g\rangle$ trap, between the two pump acts.
After each pumping process we calculate the new condensate fraction, and accordingly the new 
$T_g$, which we employ to evaluate the next pumping, and so on. 
In fig.\ \ref{fig:3} we analyze the same trap as in fig.\ \ref{fig:2}, for $k_B T_e=\omega$, and 
an initial number of trapped atoms $N=5 \times 10^4$, for initial $N_0/N=0.9$, $0.98$ and $0.99$.
One can observe that the positive effects of the reabsorption in the BAR regime allow to increase the 
condensate fraction during the pumping process, up to almost complete condensation. However, 
as pointed out previously, for very low $T_g$ the positive effects of BAR vanish. This explains the 
fact that the $N_0/N$ is slightly lower than $1$, and also that for the case of initial $N_0/N=0.99$ 
the condensate fraction initially decreases.

In this Letter we have extended the results presented in ref.\ \cite{BAR} for a much more general 3D situation, in which 
the excited state atoms can occupy more than one trapped state. We have shown that under the appropriate conditions, the BAR 
expansion is still valid for this more general case, and that the reabsorptions can play a positive role in the loading of the 
condensate. Such effect is a consequence of the always destructive interference between the processes which tend to lower the 
condensate fraction. The generalized treatment shows that the BAR condition presented in \cite{BAR} is not enough to guarantee 
the BAR expansion, since for more than one excited--trap level, the temperature of such trap is also important, and must 
be kept sufficiently low. Although for complexity reasons we have not analysed the situation in which the trap levels are 
distorted by the mean field provided by the atom--atom collisions, we must stress that similar analysis could be applied 
also if the atom--atom collisions modify the levels, if instead of the bare trap levels, self--consistent levels were considered, 
as in ref.\cite{zbigni00}.
Therefore, under the BAR conditions, i.e. large condensation, the reabsorption processes favour the optical pumping of atoms via 
spontaneous emission into a BEC. Such quantum effect is of fundamental interest, but can also be important experimentally to
overcome condensate losses, and to achieve of a continuously--loaded atom laser.
\acknowledgments
We acknowledge  support from Deutsche Forschungsgemeinschaft (SFB 
407), from the EU through the TMR network ERBXTCT96-0002, and from 
ESF PESC Programm BEC2000+, and discussions with  M. Kottke, K. Sengstock and W. Ertmer.

\end{document}